\documentstyle{mn}
\newif\ifAMStwofonts
\def\etal{{\it et al.~\/}}

\def\ie{{\it i.e.~\/}}

\def\lya{Ly$\alpha$~\/}
\def\tcmb{T_{CMB}}
\def\tcmb0{T_{CMB}^0}
\def\ltsima{$\; \buildrel < \over \sim \;$}
\def\simlt{\lower.5ex\hbox{\ltsima}}
\def\gtsima{$\; \buildrel > \over \sim \;$}
\def\simgt{\lower.5ex\hbox{\gtsima}}
\ifoldfss

  \ifCUPmtlplainloaded \else
    \NewTextAlphabet{textbfit} {cmbxti10} {}
    \NewTextAlphabet{textbfss} {cmssbx10} {}
    \NewMathAlphabet{mathbfit} {cmbxti10} {} % for math mode
    \NewMathAlphabet{mathbfss} {cmssbx10} {} %  "   "    "
  \fi
  \ifAMStwofonts
    \ifCUPmtlplainloaded \else
      \NewSymbolFont{upmath} {eurm10}
      \NewSymbolFont{AMSa} {msam10}
      \NewMathSymbol{\upi}     {0}{upmath}{19}
      \NewMathSymbol{\umu}     {0}{upmath}{16}
      \NewMathSymbol{\upartial}{0}{upmath}{40}
      \NewMathSymbol{\leqslant}{3}{AMSa}{36}
      \NewMathSymbol{\geqslant}{3}{AMSa}{3E}

      \let\leq=\leqslant \let\le=\leqslant
      \let\geq=\geqslant \let\ge=\geqslant
    \fi
  \fi
\fi % End of OFSS
\ifnfssone
  \newmathalphabet{\mathit}
  \addtoversion{normal}{\mathit}{cmr}{m}{it}
  \addtoversion{bold}{\mathit}{cmr}{bx}{it}

  \newmathalphabet{\mathbfit} % math mode version of \textbfit{..}
  \addtoversion{normal}{\mathbfit}{cmr}{bx}{it}
  \addtoversion{bold}{\mathbfit}{cmr}{bx}{it}
  \newmathalphabet{\mathbfss} % math mode version of \textbfss{..}
  \addtoversion{normal}{\mathbfss}{cmss}{bx}{n}
  \addtoversion{bold}{\mathbfss}{cmss}{bx}{n}
  \ifAMStwofonts
    \ifCUPmtlplainloaded \else
      \UseAMStwoboldmath
      \makeatletter
      \new@mathgroup\upmath@group
      \define@mathgroup\mv@normal\upmath@group{eur}{m}{n}
      \define@mathgroup\mv@bold\upmath@group{eur}{b}{n}
      \edef\UPM{\hexnumber\upmath@group}
      \new@mathgroup\amsa@group
      \define@mathgroup\mv@normal\amsa@group{msa}{m}{n}
      \define@mathgroup\mv@bold\amsa@group{msa}{m}{n}
      \edef\AMSa{\hexnumber\amsa@group}
      \makeatother
      \mathchardef\upi="0\UPM19
      \mathchardef\umu="0\UPM16
      \mathchardef\upartial="0\UPM40
      \mathchardef\leqslant="3\AMSa36
      \mathchardef\geqslant="3\AMSa3E

      \let\leq=\leqslant \let\le=\leqslant
      \let\geq=\geqslant \let\ge=\geqslant
    \fi
  \fi
\fi % End of NFSS release 1

\ifnfsstwo
  \DeclareMathAlphabet{\mathbfit}{OT1}{cmr}{bx}{it}
  \SetMathAlphabet\mathbfit{bold}{OT1}{cmr}{bx}{it}
  \DeclareMathAlphabet{\mathbfss}{OT1}{cmss}{bx}{n}
  \SetMathAlphabet\mathbfss{bold}{OT1}{cmss}{bx}{n}
  \ifAMStwofonts
    \ifCUPmtlplainloaded \else
      \DeclareSymbolFont{UPM}{U}{eur}{m}{n}
      \SetSymbolFont{UPM}{bold}{U}{eur}{b}{n}
      \DeclareSymbolFont{AMSa}{U}{msa}{m}{n}
      \DeclareMathSymbol{\upi}{0}{UPM}{"19}
      \DeclareMathSymbol{\umu}{0}{UPM}{"16}
      \DeclareMathSymbol{\upartial}{0}{UPM}{"40}
      \DeclareMathSymbol{\leqslant}{3}{AMSa}{"36}
      \DeclareMathSymbol{\geqslant}{3}{AMSa}{"3E}

      \let\leq=\leqslant \let\le=\leqslant
      \let\geq=\geqslant \let\ge=\geqslant
    \fi
  \fi
\fi % End of NFSS release 2
\ifCUPmtlplainloaded \else
  \ifAMStwofonts \else % If no AMS fonts
    \def\upi{\pi}
    \def\umu{\mu}
    \def\upartial{\partial}
  \fi
\fi
\title{Limits on Dust and Metallicity Evolution of \lya
Forest Clouds from COBE}
\author[A. Ferrara et al.]
       {Andrea Ferrara$^1$, Biman Nath$^2$, Shiv K. Sethi$^3$ \& Yuri
	   Shchekinov$^{1,4}$\\
	   $^1$Osservatorio Astrofisico di Arcetri 50125 Firenze, Italy
	   \\ E--mail: ferrara@arcetri.astro.it\\
	   $^2$Raman Research Institute, Bangalore --560080, India
	   \\ E--mail: biman@rri.ernet.in \\
	   $^3$ Institut d'Astrophysique, 75014 Paris, France
	   \\ E--mail: sethi@iap.fr\\
	   $^4$Department of Physics, Rostov State University,
	   344090 Rostov on Don, Russia 
	   \\ E--mail:  yus@rsuss1.rnd.runnet.ru} 
\date{}
\pagerange{\pageref{firstpage}--\pageref{lastpage}}
\pubyear{1997}
\begin{document}
\maketitle
\label{firstpage}
\begin{abstract}

We consider possible observational consequences of dust and metals 
in \lya forest clouds.
We relate the dust content, $\Omega_d^{Lya}$, to the metal evolution of the
absorbers and assume that dust is heated by the ultraviolet background radiation
and by the CMB.  We find that the dust temperature deviates from $T_{CMB}$
by at most 10\% at redshift $z=0$. 
The \lya cloud dust opacity to redshift $\sim 5$ sources around the
observed wavelength $\lambda_0 \sim 1 \mu$m is $\sim 0.13$, and could
affect observations of the distant universe in that band.
The expected CMB spectral distortions due to
high-$z$ dust in \lya clouds
is $\sim 1.25-10$ smaller than the current COBE upper limit, depending on
the metallicity evolution of the clouds. 
If \lya clouds are clustered, the corresponding CMB anisotropy due to dust
is $\sim 10^{-1}$ on angular scales $\theta \simlt 10''$ at frequencies probed
by various future/ongoing FIR missions,  which makes these fluctuations
potentially detectable in the near future.   
Emission from CII fine-structure transitions
could considerably contribute to submm range of the FIR background radiation. 
Depending 
on the ionization of carbon and on the density of metal enriched regions, 
this contribution can be comparable with the observed residual flux at 
$\lambda\approx 0.15$ mm, after CMB subtraction. We argue that constraints 
on metal evolution versus redshift 
can be obtained from the observed flux in that range.

\end{abstract}
\section{Introduction}
Background emission from high redshift dust has been shown to be 
a good tracer of energy release in the early universe, since 
dust optical properties differ considerably from a black body 
(Rowan - Robinson \etal 1979, Zinchenko, 1979, Negroponte \etal 1981, Bond
\etal 1986, Naselskii \& Novikov 1989, Bond \etal 1991, 
Loeb \& Haiman 1997). Direct indications of the existence
of dust at high redshift come from the reddening of
background quasars (Ostriker \& Heisler 1984, Fall \& Pei 1989).
Indirect evidences of dust in damped \lya systems have been
obtained from the relative gas-phase abundances of Zn and Cr (Pettini 
\etal 1997). Fall \etal (1996) have calculated the cosmic infrared background 
from dust in damped \lya systems, and found good agreement 
with FIR background ($\nu I_\nu \approx 10^{-5}$ erg cm$^{-2}$ s$^{-1}$ 
sr$^{-1}$ at 200 - 400
$\mu$m) deduced from {\it COBE}/FIRAS data (Puget \etal  1996, Guiderdoni
\etal 1997, Fixen \etal 1998),
which also seem to imply the presence of dust. 
Recent detection of heavy elements, such as carbon and silicon (Tytler \etal
1995, Songaila \& Cowie 1996) in \lya clouds at redshift $z\sim
3$ can potentially indicate that dust exists also in the \lya forest: it is 
quite natural to assume that dust grains are associated with  heavy elements. 
Dust in \lya forest clouds would be relevant to the  
understanding of the origin of \lya clouds and their association to
Pop~III objects (Ciardi \& Ferrara, 1997), the enrichment of 
intergalactic medium with metals (Gnedin \& Ostriker 1997), and the thermal 
history of \lya clouds (Nath \etal 1997). Here we 
assume that \lya clouds contain dust, and calculate its contribution 
to the submillimeter background. 
Our model can then be used to set  limits 
on the epoch of dust formation (possibly related to the IGM metal
enrichment era) by comparing the calculated dust 
background flux with {\it COBE}/FIRAS data. The results are independent of 
the star formation history in the universe and weakly dependent on the
specific form of the heating radiation. We first (\S~2) calculate 
the dust temperature and opacity (\S~3); 
in \S~4 we calculate the expected CMB distortions by dust
emission and \S~5 contains the estimated emission from PAH and metals in the gas
phase. \S~6 is devoted to the study of CMB  
anisotropies induced by dust in \lya clouds clustered on large scales; finally,
\S~7 summarizes the results.
\section{\lya Cloud Dust Temperature}
A dust grain in a \lya cloud is exposed to the UV-optical background
radiation (UVB) provided by high redshift QSOs (and possibly star 
forming galaxies, Loeb \& Haiman 1997, neglected here) and to 
the cosmic microwave background (CMB).
Hence, the grain will be heated to an equilibrium temperature, $T$, 
and will itself radiate. The intensity of radiation, $I_\nu(t)$,
observed at epoch $t$  is the sum of the contribution from CMB and 
UVB, both attenuated by dust, and from dust emission. This is
obtained by solving the radiative transfer equation
in an expanding universe (Rowan-Robinson \etal 1979, Negroponte \etal 
1981): 
\begin{eqnarray}
\label{inu}
I_\nu(t)=\Big\{\pi B_\nu[T_{CMB} (t)] + \pi J_\nu(t)\Big\} 
e^{-\tau_\nu(t_i,t)}
\nonumber\\
+ \int_{t_i}^{t} \pi B_\nu\left[T(s){R(s)/R(t)}\right] e^{-\tau_\nu(s,t)}
d\tau_\nu(s,t)
\end{eqnarray}
where $\pi B_\nu$ is the Planck function in units
of ergs cm$^{-2}$ s$^{-1}$ Hz$^{-1}$, $T_{CMB}$ is the cosmic microwave
background temperature, and $4\pi J_\nu$ is the intensity of the UVB
in the same units as  $\pi B_\nu$; $R$ is the scale factor, 
$t_i$ ($z_i$) is the maximum (minimum) between the 
time  (redshift) of dust formation and IGM reionization due to the UVB; 
$\tau_\nu(t_1,t_2)$
is the optical depth at epoch $t_1$ to a source emitting at epoch $t_2$.

The dust temperature at redshift $z$ is given by the detailed balance
equation:
\begin{equation}
\int_0^\infty d\nu Q_{abs}(\nu,a)\pi B_\nu[T(z)] =
\int_0^\infty d\nu Q_{abs}(\nu,a)I_\nu(z),
\label{detbal0}
\end{equation}
or
\begin{eqnarray}
\int_0^\infty d\nu Q_{abs}(\nu,a)\pi\left[B_\nu[T(z)]-B_\nu[T_{CMB}(z)]
\right]e^{-\tau_\nu(z,z_i)}\simeq\nonumber\\
\noindent\int_0^\infty d\nu Q_{abs}(\nu,a)\pi J_\nu(z) 
e^{-\tau_\nu(z,z_i)}\equiv{\cal H}(z,a).
\label{detbal}
\end{eqnarray}
$Q_{abs}(\nu,a)$ is the absorption coefficient 
for a grain of radius $a$. We have neglected in eq. \ref{detbal} the effect of 
re-radiation from grains:
this is a good approximation as long as $\tau \ll 1$. We will
show in the next Section that indeed this is the case for the
present problem. For the same reason the attenuation both of
the CMB and UVB fluxes can be neglected.
In this limit the  formal solution of eq. \ref{detbal} is 
\begin{equation}
T(z)=\left[T^6_{CMB}(z) + \Theta {{\cal H}(z,a)\over a}\right]^{1/6},
\label{tdust}  
\end{equation}
where $\Theta = c^4 h^5/2\pi Q_{abs}^0 k_B^6 \Gamma(6)\zeta(6)$. In deriving 
eq. \ref{tdust} we have used the following approximation for the absorption
coefficient at infrared wavelengths: $Q_{abs}(\nu,a)=  Q_{abs}^0 a
\lambda^{-2}$ with $Q_{abs}^0=1.182\times 10^{-2}$~cm; this is an 
excellent fit to the results of Draine \& Lee (1984) for the
astronomical silicate for $\lambda \simgt 20\mu$m; 
$\zeta$ is the Riemann function and other symbols appearing in $\Theta$ 
have usual meaning. Thus, the evolution of the dust temperature excess over
$T_{CMB}$ is governed by the UVB energy flux absorbed by the grain, 
${\cal H}(z,a)$. 
In order to determine this quantity we have adopted the Haardt \& Madau 
(1996,HM) 
UVB spectrum, and convolved it with the $Q_{abs}(\nu,a)$ as calculated from
optical properties of Draine \& Lee (1984) astronomical silicates for a
given value of $a$, in the wavelength range 
$4.3$\AA $\le \lambda \le 48 \mu$m.  
This choice tends to underestimate the true UVB flux
for two reasons: (i) possible contributions from galaxies in addition to
quasars have not been included; 
(ii) the power-law
approximation to the typical UV-optical spectrum of a 
quasar does not account for line emission 
and other features, such as the well known blue-bump.
Thus, eq. \ref{tdust} represents a lower limit to $T(z)$. 
In addition, if a population of very small grains and/or PAHs does 
exist in \lya clouds the grain temperature might fluctuate 
stochastically. 
The equilibrium approximation breaks down when the
temperature fluctuation $\Delta T$ is  
$(\langle\Delta T^2\rangle)^{1/2}/T \sim 1$:
according to Draine \& Anderson (1985) this should occur for 
$a \simlt 20$~nm,
both for silicate and graphite grains. In general this non-equilibrium  
effect  mimics a higher equilibrium temperature. 

We have solved eq. \ref{tdust} numerically, and the following 
analytical expression approximates the numerical  results within less
than 5\% on the entire redshift range for which the UVB data exist, 
$0\le z\le 6$:
\begin{equation}
{T(z)-T_{CMB}(z)\over T_{CMB}(z)}=
{\Delta T\over T_{CMB}}\approx 0.54~e^{-1.36 z} a_{nm}^{-0.37}, 
\label{dtont}
\end{equation}
where $a_{nm}=a/$nm. The previous formula demonstrates that the dust
temperature is governed by the CMB at high redshift ($\Delta T/T_{CMB}[z>3]  
\simlt 0.1\%$ for a $100$~nm grain, and therefore the presence of the
UVB is completely negligible at these earlier epochs), and that larger grains 
tend to be slightly colder, since they are more efficient emitters.
\section{Dust Opacity}
The dust optical depth at redshift  $z_1$ to a source emitting 
at redshift $z_2$ is (Rowan - Robinson \etal 1979)
\begin{equation}
\label{tau}
\tau_{\nu}(z_1,z_2)= c \int_{z_1}^{z_2} \rho_c \Omega_{d}^{Ly\alpha}(z)\kappa(\nu,a)
\Bigl| {dt\over dz}\Bigl| dz,
\end{equation}
where $\vert dt/dz \vert = H_0^{-1}(1+z)^{-5/2}$ (we assume a cosmology
with $\Omega=1$, and $h=0.65$, wherever not differently specified);
$\Omega_{d}^{Ly\alpha}$ is the mass density of dust in \lya clouds    
in units of the cosmological critical density, $\rho_c$, 
and $\kappa = 3
Q_{abs}/4a\delta$, with $\delta\sim 3$~g~cm$^{-3}$ being the grain mass 
density, is the grain absorption cross section per gram.
Introducing the dust-to-gas ratio,
${\cal D}$, we can express the dust density in terms of the \lya cloud baryon 
density as $\Omega_{d}^{Ly\alpha}= {\cal D} \Omega_{b}^{Ly\alpha}$.
Next we assume that ${\cal D}$ has a power-law dependence on the gas 
metallicity (in solar units): ${\cal D} = {\cal D}_\odot Z^p$, 
where ${\cal D}_\odot$ 
gives the appropriate normalization to the galactic value. 
The precise determination of ${\cal D}_\odot$ still suffers from uncertainties
related to the amount of cold ($T\le 30$~K) dust; reasonable estimates of
the latter quantity yield a value  ${\cal D}_\odot \approx 6.7\times 10^{-3}$,
which we use here (Young \& Scoville 1991). The index $p$ is usually taken to be
equal
to unity (Pei \& Fall 1995, Fall \etal 1996, Eales \& Edmunds 1996), 
implying a linear relation between ${\cal D}$ and $Z$. However, an 
increasing number of studies (Issa \etal 1990, Schmidt \& Boller 1993, 
Lisenfeld
\& Ferrara 1998) find that such relation is nonlinear. For example, 
Lisenfeld \&
Ferrara (1997) find $p=1.85^{+1.1}_{-0.5}$ for a sample of dwarf galaxies
which might be more representative of metal poor \lya clouds. 
%To account for this effect we study the cases $p= 1,2$.   
Finally, we need to specify the metallicity evolution of \lya clouds, $Z(z)$.
A handful of metal absorption line detections in the \lya forest with $N_{HI}
\simgt 10^{14}$~cm$^{-2}$ have now been claimed (Cowie \etal 1995, 
Tytler \etal
1995, Songaila \& Cowie 1996); nevertheless, our understanding of this issue 
remains very poor. We assume that the metallicity in these objects 
varies as 
\begin{equation}
Z(z)=Z_o\left({1+z_m\over 1+z}\right)^q, ~~~q \ge 0
\end{equation}
and use observations to fix $Z_o=Z(z_m)\simeq 10^{-2}$ at the reference 
redshift $z_m =3$. Note that $q=0$ corresponds to no-evolution, whereas
$q=3.3$ gives a present solar metallicity. 
%We will explore the two cases $q=0$
%and $q=2$ and we will refer to them as the ``no-evolution'' and ``strong
%evolution'' cases, respectively. 
Using the above prescriptions, we can write
\begin{equation}
\Omega_d^{Ly\alpha}(z)={\cal D}_\odot Z_o^p \left({1+z_m\over 1+z}\right)^
{pq}\Omega_b^{Ly\alpha};
\label{omegad}
\end{equation}
we assume $\Omega_b^{Ly\alpha} \simeq \Omega_b$ in agreement with  
recent hydrodynamical simulations (Miralda-Escud\'e \etal 1996). 
Substituting the above relation into eq. \ref{tau}, we derive 
the behavior of $\tau_{\lambda_0}(0,z)$ (shown in 
Fig. 1) as a function of the present wavelength, $\lambda_0$
for different values of $z$ and $q$. From Fig. 1 we deduce that: 
(i) the opacity in the submillimiter
range is smaller than a few percent at all redshifts $z < 20$;
(ii) the opacity to UVB photons is always $\le 0.1$ 
if $z < 6$, the epoch for which a HM-type UVB is assumed to exist. 
However, the \lya cloud dust opacity to redshift $\sim 5$ sources around 
$\lambda_0 \sim 1 \mu$m is as high as $\sim 0.13$, and could considerably 
affect observations of the distant universe in that band.
\begin{figure}
\vspace{8cm}
\caption{\footnotesize{Dust opacity as a function of the present
observation wavelength for $p=1$, $a=100$~nm. The curves refer to 
different values of the evolution parameter $q=0$ (solid lines),$1$ (dashed) 
and 
redshift $z=5,10,20$ from the lowermost to the uppermost curves, respectively.}}
\end{figure} 

%{\tt 1) dust in \lya clouds 
%being assumed to be more homogeneously distributed than damped \lya 
%systems, produces predominantly smooth reddening of background quasars, 
%without significant obscuration and bias in QSOs counts as damped 
%\lya dust does; 

%2) In his early paper Shaver (1987, 3rd IAP Astrop Meeting ``High 
%redshift and primeval galaxies'') discusses redshift-dependent reddening 
%of quasars, and shows as an upper limit $A_v\simeq 0.1$. I think there must 
%be more recent papers on this issue, but unfortunately I do not have here 
%enough available literature to look at. In our Fig. 1 $\tau(z=5)$ at 
%$\sim 0.5-0.7~\mu$m is close to this limit (and even higher, I think, for 
%strong evolution case $q\sim 3$). This can give us an additional 
%observational restriction on dust in \lya clous.} 

\section{CMB Distortion by Dust}
In this Section we calculate the expected distortions of the CMB spectrum 
caused by dust emission with the temperature evolution eq. \ref{dtont}. 
The observed intensity at the present epoch in the microwave range 
(eq. \ref{inu}) is
\begin{eqnarray}
\label{inu0}
I_{\nu_0}(z=0)=\Big\{\pi B_{\nu_0}[T_{CMB}(z)/(1+z)]\Big\}
[1-\tau_{\nu_0}(0,z_i)]
\nonumber\\
\noindent + \int_{0}^{z_i} \pi B_{\nu_0}\left[T(z)/(1+z)\right] 
d\tau_{\nu_0}
\end{eqnarray}
where $\nu=\nu_0(1+z)$ ($\nu_0$ is the photon frequency at $z=0$).
We have used the strong inequality $\tau^{IR}_{\nu_0}\ll 1$, and in the third 
term
second order factors in the product $\Omega_d^{Lya}$ have been neglected.
We now compare the total intensity in eq. \ref{inu0} with 
that of CMB one.  To parametrize the deviation of the CMB spectrum at $z=0$
from a black--body, we introduce the $y$--parameter  
defined as
\begin{equation}
y(z_i)\equiv {1\over 4}\left[{\int_0^\infty d\nu_0 I_{\nu_0}(z_i) \over \int_0^\infty 
d\nu_0 \pi B_{\nu_0}\left[T_{CMB}(z_i)/(1+z_i)\right]}-1\right];
\label{yc}
\end{equation}
The experimental results from COBE have set the upper limit,  
$y_C = 1.5\times 10^{-5}$ 
at the 95\% confidence level (Fixsen \etal 1996).
Using eqs. \ref{dtont}, \ref{tau} and \ref{inu0} we obtain

\begin{equation}
y(z_i) \simeq {\cal A}\int_0^{z_i} dz H_0\Bigl| {dt\over dz}\Bigr| 
{(1+z)^{6-pq}e^{-1.36z}},
\label{yanal}
\end{equation}
where ${\cal A}={\cal D}_\odot Z_o^p 
\left(1+z_m\right)^{pq}\Omega_b^{Ly\alpha}
a_{nm}^{-0.37}h$.
Note that $y$ depends strongly on
$z_i$ as long as $\Delta T/T_{CMB}$ is relatively large, but only linearly 
on $\Omega_b^{Ly\alpha}$, thus making the result robust
against evolution of cloud baryon content.

Fig. 2 shows graphically the result given in eq. \ref{yanal} for
$\Omega_b^{Ly\alpha}=0.02$ and $a_{nm}=100$ 
for different values of $p$ and $q$ as a function
of $z_i$.  For $p=1$, $y(z_i)$ is falls below $y_C$ by 
a factor 1.25-10 ($z_i \simgt 3$). Higher values of $q$ (\ie stronger
metallicity evolution) produce larger values of $y$ since more dust is
present at redshifts where $\Delta T/T_{CMB}$ is larger. 
For similar reasons, the case with $p=2$ generally yields lower values
of $y$. As a comparison we have also plotted results for two different
cosmologies, \ie a flat model with $\Omega_\Lambda=0.4, h=0.65$ and an 
open model with $\Omega=0.2, h=0.65$, for $p=1, q=0$. These two models 
give essentially the same curve for the $y$-parameter, which in turn is 
lower than the standard case $\Omega=1$. 
\begin{figure}
\vspace{8cm}
\caption{\footnotesize{The equivalent $y$-parameter versus $z_i=min[$
Reionization, Dust formation redshift$]$ produced by dust ($a=100$nm) in
different evolutionary scenarios: $p=1$ (solid lines), and
$p=2$ (dotted lines) for $q=0,~1,~2,~3.3$; $\Omega_b^{Ly\alpha}=0.02$.
The {\it dashed}, overlapped lines refer to the cosmologies 
$\Omega_\Lambda=0.4, h=0.65$ and $\Omega=0.2, h=0.65$, for $p=1, q=0$.}}
\end{figure}

\section{FIR Emission from PAHs and metals}

%............. PAH etc......
Small grains ($a\sim 10-100$ \AA) and PAH molecules ($a\sim 5-30$ \AA) are 
numerous in the local ISM. They are traced by a strong excess of 
emission around 10$\mu$m (Puget etal 1985) which arises either due to 
temperature fluctuations of small grains, or characteristic IR emission of PAH 
molecules. We assume that physical conditions in \lya clouds allow small 
grains to exist there. Although the origin of small grains is still far from 
being clearly understood, we note that most efficient destruction 
mechanism (ionizing UV radiation) is much weaker (by at least 2.5 orders 
of magnitude) in \lya clouds than in the local ISM. Following this idea, we 
estimate possible observational evidences for small grains and PAH 
molecules in \lya clouds. 

The total energy emitted by a 100-atom PAH 
in a band $\Delta \lambda=0.04~\mu$m around the restframe wavelength
$\lambda_r=3.3~\mu$m can be estimated as 
$E_{\lambda_r}\simeq 4.2$ eV (Puget \& L\'{e}ger, 1989) per absorbed UV 
photon ($E_{UV}\sim 10$eV).  
The volume emissivity in this band is then
\begin{equation}
{\cal E}_\nu={E_{\lambda_r}\over \Delta \lambda}\dot N_p^{PAH}.
\end{equation}
$\dot N_p^{PAH}$ is the rate of UV photons absorbed by PAHs per unit volume 
\begin{equation}
\dot N_p^{PAH}=\Phi_{UV}\sigma_{UV}^{PAH} \Delta n_g, 
\end{equation}
where $\Phi_{UV}$ is the incident UV photon flux, 
and $\sigma_{UV}^{PAH}$ is the absorption cross section. L\'{e}ger \etal 
(1989) give $\sigma_{UV}^{PAH}\simeq 2\times 10^{-17}~N_C$ cm$^2$, 
with $N_C\approx 30$ the number of carbons in a PAH of $a\sim 10$ \AA. 
To estimate the number density of PAHs, $\Delta n_g$, we assume that the MRN
(Mathis \etal 1977) grain size distribution extends well into 
the domain of molecular sizes, following Tielens (1990): 
\begin{equation}
{dn_g\over da}=C\left({a_m\over a}\right)^{3.5},
\end{equation}
where $a_m$ is the distribution upper radius cut-off; $C$ is a constant 
to be determined from 
\begin{equation}
\int\limits_{\rm sizes} {4\over 3}\pi \delta a^3 {dn_g\over da} da
={\cal D}\Omega_b^{Ly\alpha}\rho_c.
\end{equation}
This gives 
\begin{equation}
\Delta n_g(a)= {3\over 2}{\Omega_b^{Ly\alpha}\rho_c\over \pi \delta a_m^3}
\left({a_m\over a}\right)^{3.5}\Delta a.
\end{equation}
We identify PAH molecules with grains of $a\sim 10$\AA~which contain about 
100 atoms (about 70 hydrogens and 30 carbons, see Puget \& L\'{e}ger, 
1989), and estimate their number density as 
\begin{equation}
\Delta n_g=5.5\times 10^{-9}{\cal D}_0Z_o^p\left({1+z_m\over 1+z}\right)^q
(1+z)^3\Omega_b^{Ly\alpha}\Delta a, 
\end{equation}
with $\Delta a\simeq a\approx 10^{-7}$ cm; here we have taken $a_m=0.1~\mu$m. 
Assuming the photon flux at $E_{UV}\sim 10$ eV to be 
$\Phi_{UV}\sim 3\times 10^7$ phot s$^{-1}$ cm$^{-2}$ -- a factor 3 higher 
than the HM value at $z\approx 2-3$ to account for possible
contributions from normal galaxies -- we then obtain    
%\begin{equation}
%{\cal E}_\nu=3.6\times 10^{-40}{\cal D}_0 Z_o^p
%\left({1+z_m\over 1+z}\right)^q(1+z)^3\Omega_b^{Ly\alpha}\Delta a, 
%\end{equation}
%which being substituted in 
\begin{eqnarray}
\nu_0 I_{\nu_0}={c\over 4\pi}\int_0^\infty dz\left|{dt\over dz}\right|
{\cal E}_{(1+z)\nu_0}
\nonumber\\
\simeq 2\times 10^{-12} {\rm erg~ s^{-1}~ cm^{-2}~ sr^{-1} }. 
\end{eqnarray}
Similar values are found for other PAHs bands, \ie $\lambda_r=6.2,~7.7,~ 8.6,~
11.3~\mu$m 
Thus the characteristic emission from PAHs
in \lya clouds is much weaker than the FIR background deduced by FIRAS,
and shown in Fig. 3.

The contribution of temperature fluctuations of small grains ($a< 100$ \AA) 
to the CMB distorsion can also be shown to be unimportant. This can be seen 
from a comparison of the characteristic time between two subsequent absorptions 
of UV photons, 
\begin{equation}
t_{abs}\approx 4\times 10^5\left({100~{\rm \AA} \over a}\right)^3~\rm s,
\end{equation}
and the characteristic cooling time of a grain,
\begin{equation}
t_{cool}\approx {10^4\over T_0}~ \rm s, 
\end{equation}
where the maximum temperature of a grain $T_0$ after absorption of a photon 
with energy $h\nu\approx 20$ eV is $T_0\approx 50(100~{\rm \AA}/a)^{3/4}$ (Aanestad, 
1989). This gives 
\begin{equation}
{t_{cool}\over t_{abs}}\simeq 5\times 10^{-4}\left({100~{\rm \AA} \over 
a}\right)^{-15/4}\ll 1.
\end{equation}
%One can show also that the luminosity of a Ly$\alpha$ cloud in PAHs bands 
%is well below the observational limits 
%
%\begin{eqnarray}
%L_\nu\sim 4\times 10^{47}{\cal D}_0 Z_o^p\left({1+z_m\over 1+z}\right)^q
%(1+z)^3\Omega_b^{Ly\alpha}\Delta a~
%~~~~~\nonumber\\
%\simlt 8\times 10^{34}~{\rm erg~s^{-1}},
%\end{eqnarray}
%in $\lambda_0=3.3~\mu$m band, for a cloud of 100 kpc radius, and 100 
%overdensity. 

The FIR background emission from ions in gas phase, particularly from CII 
(at rest-frame wavelength $\lambda_r=158~\mu$m) and FeII (rest-frame wavelengths 
$\lambda_r=26,~35~\mu$m) can potentially trace heavy elements in \lya clouds. 
The CII emissivity per unit volume due to collisions is 
(see Wolfire \etal 1995) 
\begin{equation}
{\cal E}_\nu=5\times 10^{-25}Z_o[C] x_{_{CII}}
T_4^{-0.18}n_e^2(\Delta \nu_D)^{-1},
\end{equation}
where $\Delta \nu_D$ is the Doppler width, $[C]$ the abundance of 
carbon, $x_{_{CII}}$, the CII ionization 
fraction. With the above assumptions and for  solar 
carbon abundance, $[C]=[C]_\odot$, we obtain 

\begin{eqnarray}
\nu_0I_{\nu_0}={c\over 4\pi } \nu_0\int_0^{z_i}dz \left|{dt\over dz}\right|
{\cal E}_{(1+z)\nu_0}
\nonumber\\
~~~~~\simeq 
10^{-13}h^3 f \Delta^2 \left({\Omega_b^{Ly\alpha}\over 0.02}\right)^2x_{_{CII}}
(1+z_m)^q
~~~~\nonumber\\ 
\times\left({\nu_0\over \nu_r}\right)^{q-1/2}\Theta[\nu_0(1+z_i)-\nu_r],
\label{cflux}
\end{eqnarray}
%\begin{equation}
%{\cal E}_\nu\simeq 1.5\times 10^{-30} Z_oT_4^{-0.68}h^4(1+z)^6
%(\Omega_b^{Ly\alpha})^2\nu^{-1},
%\end{equation}
%and for $Z_o=10^{-2}$, $\Omega_b^{Ly\alpha}=0.02$ and $z=10$ we get 
%$\nu I_nu\sim 10^{-6}$ erg s$^{-1}$ cm$^{-2}$ sr$^{-1}$ at $\nu=1.58$ mm. 
where $\Delta$ is the overdensity of \lya clouds, and $f$ is the filling 
factor of denser regions.
Fig. 3 shows the expected emission from CII ions compared with the FIR
background. We are plotting for completness three different estimates
of the background taken from Puget \etal (1996), Guiderdoni \etal (1997), 
and Fixsen \etal (1998).
Such emission in this case is  more model dependent than the CMB distortion
produced by dust particles. Besides uncertainties connected with abundance 
and ionization of CII, the density of \lya cloud gas enters  as an 
additional free parameter due to the collisional nature of emission from
CII fine-structure lines. On average, the density contrast of \lya
clouds is thought not to exceed 10-30. However, the regions 
containing metals and dust could be connected with debries of supershells produced 
by strong explosions during active star formation phases in young galaxies, and 
can be much more dense. The gas density in such regions can possibly cover 
a wide range up to galactic values, $n\sim 0.1$ 
cm$^{-3}$. To illustrate the dependence of the FIR background on gas density in 
metal-rich regions of \lya clouds, we present in Fig. 3 the flux calculated 
from eq. \ref{cflux} for density contrasts $\Delta=10^3,~3\times 10^3$, and 
$10^4$ which correspond at $z=3$ to a number density $n\simeq 0.005,~0.015$, 
and 0.05 cm$^{-3}$, respectively. The intensity of CII emission is 
plotted for $x_{_{CII}}=1$, and scales as $\propto x_{_{CII}}$. 

\begin{figure}
\vspace{8cm}
\caption{\footnotesize{The FIR background and submm emission from CII ions in 
\lya clouds. Solid lines show different estimates of the {\it COBE}/FIRAS data 
by Puget \etal (1996) (labelled P), Guiderdoni \etal (1997) (G), and Fixen 
\etal (1998) (F). 
Broken lines represent CII emission for different 
metallicity evolution: $q=3.3,~ 2.0,~1.0,~0.0$, and \lya cloud overdensity 
$f^{1/2}\Delta=10^4$ (dot-dashed), $f^{1/2}\Delta=3\times 10^3$ 
(long dot-dashed), $f^{1/2}\Delta=10^3$ (dashed); 
fractional ionization $x_{_{CII}}=1$, $z_m=3$, 
$z_i=10$. 
}}
\end{figure}
Fig. 3 allows us to draw qualitative conclusions about 
the evolution of metals in \lya clouds. It is seen clearly that only  
strong evolutionary scenarios ($q>1$) can avoid producing 
an excess in the submm range of the backround radiation. In particular, 
for $x_{_{CII}}=0.01$ models with high overdensity, $f^{1/2}\Delta=10^4$, 
produce 
an intensity below the {\it COBE}/FIRAS data at submm wavelengths only when $q\geq 2$. 
In general, this requirement is satisfied by all models with 
$f^{1/2}\Delta x_{_{CII}}^{1/2}\leq 10^3$. 
For realistic values $f<\Delta^{-1}$, the CII emission can exceed the 
{\it COBE}/FIRAS flux in the submm range only for overdensities 
$\Delta\simgt 10^6$. 

It is clear also that FeII ions give small contribution to the FIR background 
for two reasons: iron is less abundant than carbon, and in addition 
fine-structure emission of FeII falls to shorter wavelength range,  
$26~(35) \mu{\rm m}\leq\lambda_0\leq 26~(35)~(1+z_i)~\mu{\rm m}=286~(385)\mu$m 
for $z_i=10$, where the background emission is much stronger than in 
submm range. 

%             End

%............ Shiv's chapter

\section{CMB Anisotropies due to Dust Emission}

As the dust distribution is inhomogeneous, it will result in the 
anisotropy of the FIR  background radiation
caused by dust emission. The angular 
pattern of anisotropy will depend on the sizes and clustering properties of 
the Lyman-$\alpha$ clouds. If the Lyman-$\alpha$ clouds are assumed to 
be unclustered and Poisson distributed then the typical value of anisotropy
in the intensity
$ \simeq N^{-1/2}$, $N$ being the average number of clouds along any
given line of sight; observations show that $N \simeq 100$ up to $z \simeq 3$
(see e.g., Mo \etal 1993)
The angular pattern will be  a 'white noise' spectrum, i.e., 
independent of the  angle. Of course, in a realistic experiment there would be a further
smearing of this anisotropy by the finite width of the beam by a factor of 
$x_{\rm cl}/\sigma_0$, $x_{\rm cl}$ and $\sigma_0$ being the typical size of 
cloud and the beam width, respectively. However, though the clustering 
properties of Lyman-$\alpha$ clouds remain controversial, Cristiani \etal  
(1997)
have claimed a weak but non-negligible
 clustering of Lyman-$\alpha$ clouds at high
redshifts. According to Cristiani \etal  (1997), Lyman-$\alpha$ clouds below
HI column densities  $10^{13.8} \, \rm cm^{-2}$ show no clustering but clouds
of higher column density cluster at scales corresponding to a few hundred
${\rm km s}^{-1}$, with a clustering amplitude proportional to the HI column 
density. In addition, they notice a trend of increase in clustering as the 
redshift decreases. Their results, cast in the form of three-dimensional 
two-point 
correlation function,  are consistent with $\xi(r) = [r/r_0(z)]^{-\gamma}$ with
$\gamma = 1.77$ and $r_0(z) = r_0(0) \times (1+z)^{-5/3}$ from $1.7 \simlt 
z \simlt 4$ with $r_0(0) \simeq 1.5 \hbox{--}2 h^{-1} \, \rm Mpc$.
 At lower redshifts the clustering is observed to be higher than
predicted by the clustering  behaviour at high $z$ (Ulmer 1996). 
For the purposes of this
paper, we take the redshift dependence of clustering amplitude indicated by
the high redshift data up to $z=0$, bearing in mind that this might
underestimate
the clustering at low $z$. 

We follow the treatment of Bond et al. (1986) in evaluating the two point
angular correlation function, $C(\theta)$, of the anisotropies from the clustering of Lyman-$
\alpha$ clouds (for more details see Peebles 1980). For the three-dimensional
two-point correlation function given above, we obtain
\begin{equation}
C^{1/2}(\theta) = {\langle \Delta I_d^2(\theta)\rangle^{1\over 2}\over 
\langle I_d \rangle}\simeq F_{pq}(z_i) 
\left({H_0r_0(0)\over 2c}\right)^{\gamma\over 2} 
\sin^{1-\gamma\over 2}\theta, 
\label{angan}
\end{equation}
where  $\langle I_d\rangle$ is
the mean intensity of the background emission from dust. $F_{pq}(z_i)\simeq 1$, 
and
it increases slightly  as  $z_i$ is decreased and/or the product $pq$ is
increased, indicating  that the dominant signal comes from a smaller 
redshift where the clustering scale is larger. For
$\gamma\simeq 1.77$ and $H_0 r_0(0)\simeq 150 \, \rm  km s^{-1}$, we get  
$\langle \Delta I_d^2\rangle^{1/2}/\langle I_d \rangle \simeq 1.1\times 10^{-3}
\theta^{-0.4}$ for $\theta \ll 1$. At $\theta \simeq 10''$, the angle
corresponding to the typical clustering scale at high $z$,    
 $\langle \Delta I_d^2\rangle^{1/2}/\langle I_d \rangle \simeq 0.1$,  which
is comparable to what one would expect from the `white noise' fluctuations
discussed above. We recall that 
eq.~(\ref{angan}) should not be used for scales smaller
than the typical size of the cloud at high $z$. The signal from dust emission
will be significant at wavelengths  $\simeq 400 \hbox{--}1000 \, \rm \mu m$
with typical intensities $\simeq 10^{-7}\hbox{--}10^{-8} \, \rm erg \, cm^{-2}
\, sec^{-1} \, sr^{-1}$,  with a fluctuation amplitude of $0.1$ at angular 
scales $10''\hbox{--}1'$. The ongoing and future missions 
like ISO (ISO-PHOT), 
SCUBA, and FIRST operate/will operate
 in this wavelength band with sensitivities high
enough to detect this fluctuation, though it may be hard to separate the 
contribution of \lya cloud dust emission from other
extragalactic sources at high redshift.

Another potentially measurable quantity is the  predicted CMB anisotropy.
The latter is  given  by the ratio of the angular intensity variation to 
the CMB radiation at any given frequency:

\begin{eqnarray}
{\langle \Delta I_d^2(\theta)\rangle^{1/2}\over 
\langle I_{CMB} \rangle}= 
{3.55\times 10^{-8}\over 11-2pq}
Z_o^p\left({\Omega_b^{Ly\alpha}\over 0.02}\right)(1+z_m)^{pq} \times
~~~~~~\nonumber\\
\left ({\nu_0 \over 10^{11} \, \rm Hz}\right )^2
\left[(1+z_i)^{-pq+{11/2}}-1\right]
{\langle \Delta I_d^2(\theta)\rangle^{1/2}\over 
\langle I_d \rangle}. 
\label{relan}
\end{eqnarray}

Taking $z_i = 5$, $p = q = 0$, and $\Omega_b^{Ly\alpha} =0.02$, we get
$\langle \Delta I_d^2(\theta)\rangle^{1/2}/ 
\langle I_{CMB} \rangle\simeq 9.2\times 10^{-6}h^{-1} 
\langle \Delta I_d^2(\theta)\rangle^{1/2}/ \langle I_d \rangle$ at 
$\nu_0=360 \, \rm  GHz$. 
Most of the  contribution to the anisotropy comes from high redshifts (around $z = z_i$)
where most of the baryons are seen to be in Ly$\alpha$ clouds in recent
hydrodynamical simulations (Miralda-Escud\'e et al. 1996.). 
Fomalont et al. (1993) obtained  an upper bound
of $\Delta T/T_{CMB} < 3 \times 10^{-5}$ for $10''\hbox{--}100''$ 
at $\nu_0 = 8.44 \, \rm GHz$. The  anisotropies from dust emission are several
orders below this upper limit because of their $\nu_0^2$ dependence. 
The best hope to detect the anisotropy is at infrared frequencies. 
Recently, the ongoing CMB experiment Sunyaev-Zel'dovich Infrared
experiment (SuZIE) reported its first results. At $\nu = 142 \, \rm  GHz$,
they obtained an upper limit $\Delta T/T \le  2.1 \times 10^{-5}$ at
angular scales of $\simeq 1'.1$ (Church \etal 1997).
 A comparison of this result with our
estimate shows that our predictions are still an order below this 
upper limit. However, in coming years, the SuZIE instrument will be
upgraded to have additional frequency channels at $217$ and $268 \,
\rm GHz$.  As the expected signal is much higher at these frequencies,
it might be marginally detectable. Also, this anisotropy might be
an important foreground for other secondary CMB anisotropies which
could be  present at sub-arc minute scale with amplitudes of a few $\mu
\, \rm K$ (for details see Bond 1996 and  references therein).
 The future
CMB mission  Planck surveyor  will also operate at frequencies at which the 
dust emission from high $z$ Ly$\alpha$ clouds will peak (see Bouchet \etal 
1995)
With its sensitivity, the Planck 
surveyor will be able to 
detect a fluctuation  at the level of a few $\mu \, K$ at $\nu_0 \simeq 300
\, \rm GHz$ with an angular resolution of $4.5'$. From our analysis above,
the fluctuations expected from dust
emission are likely to be weaker than the sensitivity of the Planck surveyor.

\section{Conclusions}

We have considered possible observational consequences of the presence of 
dust in \lya clouds. We have used a rather general expression for the
dust density, $\rho_d(z)=\rho_c {\cal D}_0Z_o^p(1+z_m/1+z)^{pq}
\Omega_b^{Ly\alpha}(z)$, which accounts for possible nonlinearity of 
the dust-to-metals ratio versus metallicity (Lisenfeld \& Ferrara 1998),
and evolution of metallicity with $z$.
We have calculated the temperature of dust grains heated by CMB photons and
UV background radiation from quasars following Haardt \& Madau (1996), and  
assuming dust to be constituted by standard ``astronomical silicates''
(Draine \& Lee, 1984). Our main results are the following: 

1) The contribution of UV radiation to dust heating is small in
comparison with heating by CMB photons; only at $z\leq 1$ the dust
temperature fractional deviation from $T_{CMB}(z)$ reaches its maximum of 5-10\%. 

2) The \lya cloud dust opacity to redshift $\sim 5$ sources around the
observed wavelength $\lambda_0 \sim 1 \mu$m is $\sim 0.13$

3) The expected CMB spectral distortion due to high-$z$ dust in \lya clouds
is $\sim 1.25-10$ smaller than current COBE upper limit on the $y$-parameter, 
depending on the metallicity evolution. 

4) The predicted value of  anisotropies produced by dust 
clustered on scales $H_0 x_0\sim 150$ km s$^{-1}$  is   
$\langle \Delta I_d^2(\theta)\rangle^{1/2}/ 
\langle I_{d} \rangle\simeq  0.1$ on scales $\theta \simlt 10''$.
The ongoing and future missions operating at FIR wavelengths might
be able to detect such effect, thus providing   
 valuable information on 
the dust content, evolution and clustering properties of \lya clouds.

5) The fine-structure emission of CII ions can contribute 
to the submm range of the spectrum if metal enriched regions in \lya clouds are 
sufficiently dense, $n\sim 0.1$ cm$^{-3}$. If the average overdensity in 
such regions is as high as $\Delta\simgt 10^3/({f x_{_{CII}}})^{1/2}$, only 
strong metallicity evolutionary scenarios with $q>1$ are allowed.  

\vskip 1truecm

We thank the referee, Dr. F. Bouchet, for insightful comments.
SKS thanks Bruno Guiderdoni for useful information on FIR missions. 
YS acknowledges support from NATO Guest Fellowship Programme 1996 
from Italian CNR 219.29

%.......................................................................

\label{lastpage}
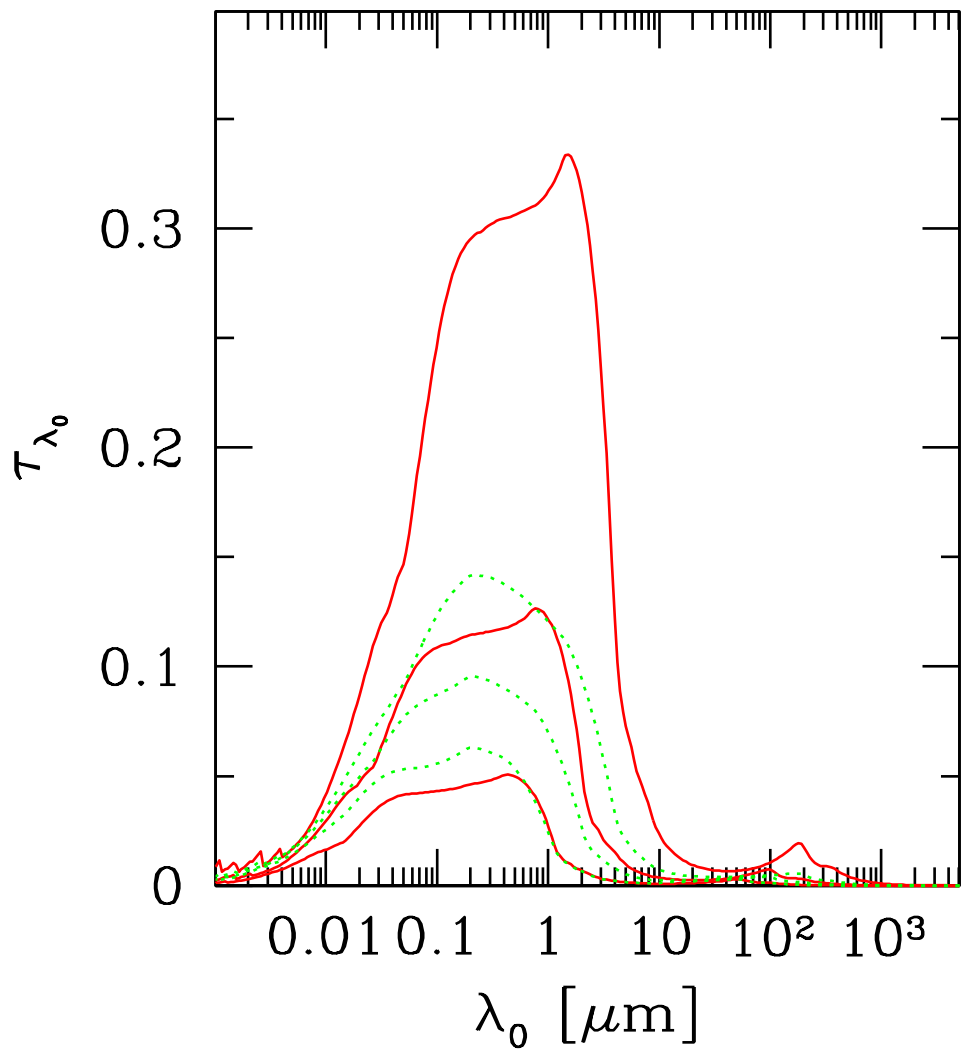
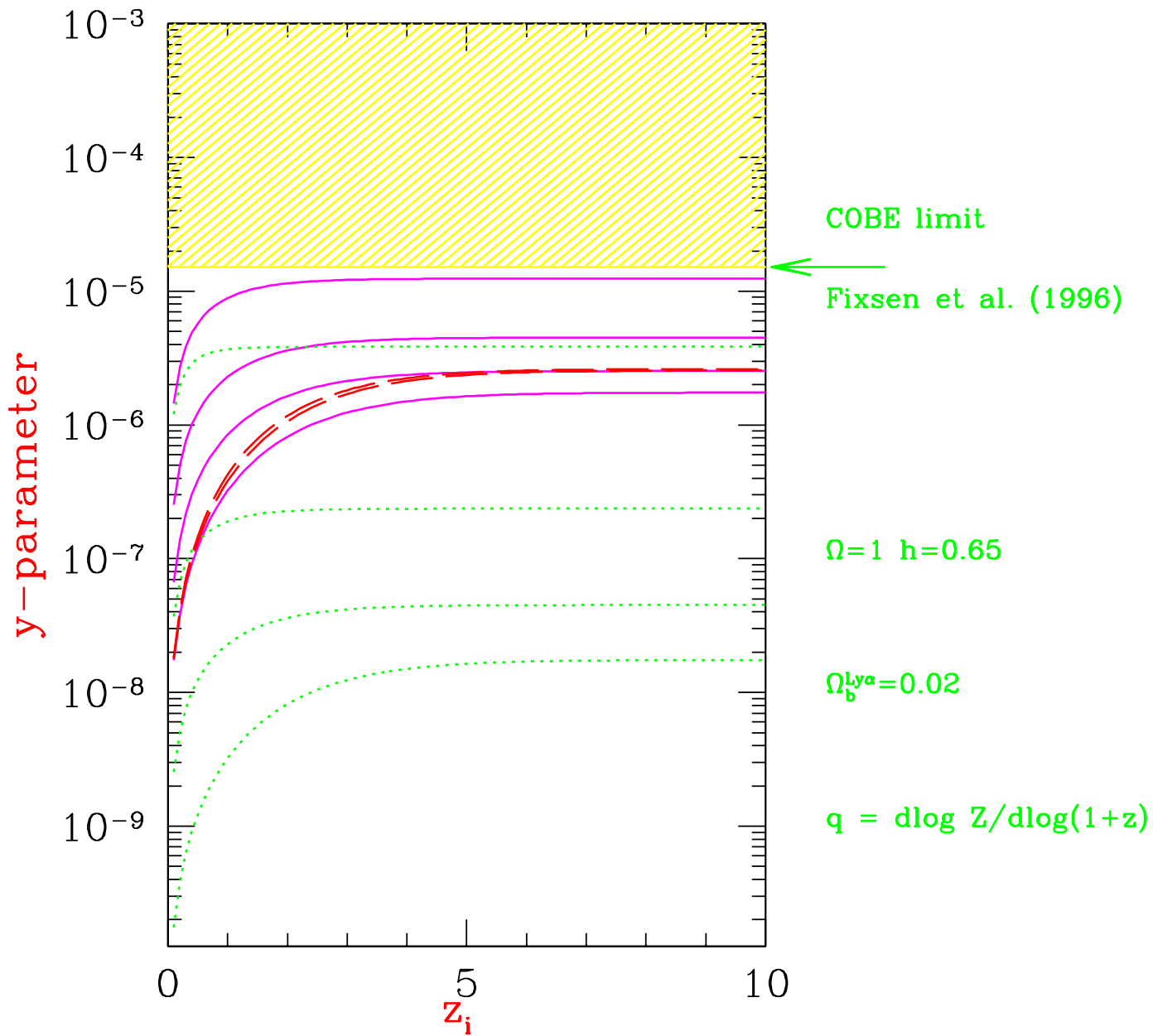
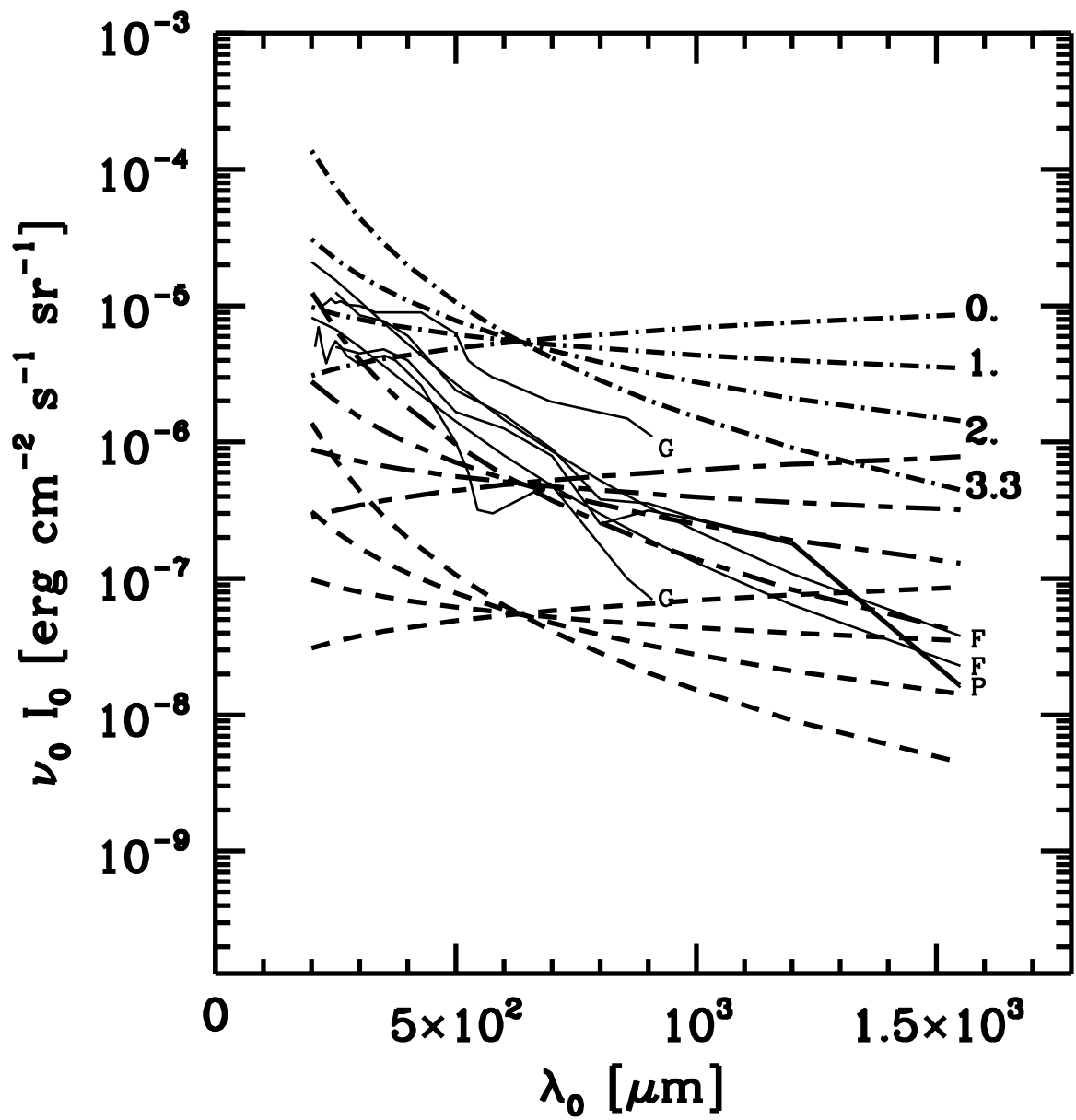
\end{document}